\documentstyle[12pt]{article}

\def\df #1. #2\par{\medbreak
  \noindent{{\tt {\bf Definition #1.}}\enspace}{\sl#2\par}%
  \ifdim\lastskip<\medskipamount \removelastskip\penalty55\medskip\fi}

\def\theorem #1. #2\par{\medbreak
  \noindent{\tt {\bf Theorem #1.}\enspace}{\sl#2\par}%
  \ifdim\lastskip<\medskipamount \removelastskip\penalty55\medskip\fi}

\def\lemma #1. #2\par{\medbreak
  \noindent{\tt {\bf Lemma #1.}\enspace}{\sl#2\par}%
  \ifdim\lastskip<\medskipamount \removelastskip\penalty55\medskip\fi}

\def\proof{\medbreak\noindent{\bf Proof}}

\def\remark{\medbreak\noindent{\bf Remark}}

\def\dsl{\raise.15ex\hbox{/}\kern-.65em\nabla}

\def\sop{$\rm SO^+(1,3)\,$}
\def\cl{{\cal C}\!\ell}

\def\one{{\bf 1}}

\def\I{{\cal I}}
\def\A{{\cal A}}
\def\M{{\cal M}}
\def\H{{\cal H}}
\def\E{{\cal E}}
\def\L{{\cal L}}
\def\V{{\cal V}}

\def\R{{\cal R}}
\def\C{{\cal C}}

\def\t{\tilde}
\def\be{\begin{equation}}
\def\ee{\end{equation}}
\def\LE{\Lambda(\E)}
\def\LkE{\Lambda^k(\E)}

\def\LCE{\Lambda_{\C}(\E)}

\def\tr{{\rm Tr}}
\def\ww{\wedge\ldots\wedge}

\def\det{{\rm det}}
\def\even{{\rm even}}
\def\odd{{\rm odd}}

\def\spin{{\rm Spin}}

\def\Spin{{\rm Spin}}
\def\com{{\rm Com}}
\def\exp{{\rm exp}}
\def\diag{{\rm diag}}
\def\s{\stackrel}

\begin{document}

\title{A tensor form of the Dirac equation }

\author{N.G.Marchuk \thanks{Research supported by the Russian Foundation
for Basic Research, grants 00-01-00224, 00-15-96073, and by the Royal Society.}}


\maketitle

Steklov Mathematical Institute, Gubkina st.8,
Moscow 117966, Russia;
nmarchuk@mi.ras.ru;
http://www.orc.ru/\~{}nmarchuk
\vskip 1cm


\begin{abstract}
We prove the following theorem: the Dirac equation for an electron
(invented by P.A.M.Dirac in 1928) can be written as a linear tensor equation.
An equation is called {\sl a tensor equation} if all values in it are tensors and all operations in it take tensors to tensors.
\end{abstract}

\vfill\eject

\tableofcontents



\section*{Introduction}
The Dirac equation for an electron \cite{Dirac} can be written in
several different but equivalent forms. In this paper we consider two
forms of the Dirac equation. Namely Hestenes' form of the Dirac
Equation (HDE) and the Tensor form of the Dirac Equation
(TDE).
The aim of this paper is to prove the following theorem.

\theorem 1. The Dirac equation for an electron (\ref{20}) (see
\cite{Dirac})
can be written in a form of linear tensor equation (\ref{1010}).

An equation is said to be written in a form of tensor equation
if all values
in it are tensors and all operations in it take tensors to tensors.

A column of four complex valued functions (a bispinor)
represents the wave function of the electron in the Dirac equation.
This wave function has unusual (compared to tensors)
transformation properties under Lorentz changes of coordinates.
These properties are investigated in the theory of spinors
(see
\cite{Cartan},\cite{Weyl},\cite{Chevalley},\cite{Benn},\cite{Lounesto}).

There were attempts to find tensor equations equivalent to the
Dirac equation \cite{Whittaker},\cite{Taub},\cite{Ruse}. The resulting
equations were nonlinear. This fact leads to difficulties with the
superposition principle, etc. The tensor equation under consideration in
our paper is linear.

In the first part of paper we consider the Clifford algebra $\cl(1,3)$
and in the second part we consider the exterior algebra of Minkowski
space $\LE$.
Elements of $\LE$ are covariant
antisymmetric tensors and elements of $\cl(1,3)$ are not tensors.

The tensor form of the Dirac equation (\ref{1010}) under consideration
has three sources. The first source is the
Ivanenko-Landau-K\"ahler equation (\ref{1080})
(see \cite{Ivanenko},\cite{Kahler}).
This is a tensor equation and a wave function of the electron is
represented in it by a nonhomogeneous covariant antisymmetric
tensor field with 16 complex valued components
(four times more than in the Dirac equation).

The second source is  Hestenes' form of the Dirac equation (\ref{320})
(see  \cite{Hestenes},\cite{Hestenes1}). The wave function of the electron is
represented in it by a real even element of the Clifford algebra
$\cl^\even(1,3)$
and has 8 real valued components. This equation is
equivalent to the Dirac equation (see the proof in \cite{Casanova}).
HDE contains nontensor values namely elements of
the Clifford algebra $\cl(1,3)$. Hence it is not a tensor equation.

The third source is the constructions developed
in \cite{Marchuk0},
\cite{Marchuk2},\cite{Marchuk3}, \cite{Marchuk5}.

Also we want to mention an interesting approach
to the construction of the 2D Dirac tensor
equation suggested by D.Vassiliev \cite{Vassiliev}.

\section{Part I.}
\subsection{The Dirac equation for an electron.}
Let $\E$ be Minkowski space with the metric tensor
\begin{equation}
g=\|g_{\mu\nu}\|=\|g^{\mu\nu}\|=\diag(1,-1,-1,-1),
\label{10}
\end{equation}
$x^\mu$ coordinates, and $\partial_\mu=\partial/\partial x^\mu$.
Greek indices run over $(0,1,2,3)$.
Summation convention over repeating indices is assumed.
Our system of units is such that speed of light $c$,
Plank's constant $\hbar$, and the positron charge have the value $1$.

The Dirac equation for an electron \cite{Dirac} has the form
\begin{equation}
\gamma^\mu(\partial_\mu \psi+i a_\mu \psi)+i m \psi=0,
\label{20}
\end{equation}
where $\psi=(\psi_1\,\psi_2\,\psi_3\,\psi_4)^T$
is a column of four complex valued functions of
$x=(x^0,x^1,x^2,x^3)$ ($\psi$ is the wave function of an electron),
$a_\mu=a_\mu(x)$ is a real valued covector of electromagnetic potential,
$m>0$ is a real constant (the electron mass),
and $\gamma^\mu$ are the Dirac $\gamma$-matrices
\begin{eqnarray}
\gamma^0&=&\pmatrix{1 &0 &0 & 0\cr
                  0 &1 & 0&0 \cr
                  0 &0 &-1&0 \cr
                  0 &0 &0 &-1},\quad
\gamma^1=\pmatrix{0 &0 &0 &-1\cr
                  0 &0 &-1&0 \cr
                  0 &1 &0 &0 \cr
                  1 &0 &0 &0},\label{30}\\
\gamma^2&=&\pmatrix{0 &0 &0 & i\cr
                  0 &0 &-i&0 \cr
                  0 &-i&0 &0 \cr
                  i &0 &0 &0},\quad
\gamma^3=\pmatrix{0 &0 &-1& 0\cr
                  0 &0 & 0&1 \cr
                  1 &0 &0 &0 \cr
                  0 &-1&0 &0}.\nonumber
\end{eqnarray}

The equation (\ref{20}) is invariant under the gauge transformation
\begin{equation}
\psi\to\psi\,\exp(i\lambda), \quad a_\mu\to a_\mu-\partial_\mu\lambda,
\label{70}
\end{equation}
where $\lambda=\lambda(x)$ is a smooth real valued function.

The matrices (\ref{30}) satisfy the relations
\begin{equation}
\gamma^\mu\gamma^\nu+\gamma^\nu\gamma^\mu=2 \eta^{\mu\nu}\one,
\label{40}
\end{equation}
where $\one$ is the $4\!\times\!4$-identity matrix and
$$
\|\eta^{\mu\nu}\|=\diag(1,-1,-1,-1).
$$

\medskip
Note that if $4\!\times\!4$-matrix $T$ is invertible, then the matrices
$\acute{\gamma}^\mu=T^{-1}\gamma^\mu T$ also satisfy (\ref{40}).
\medskip

Let us denote
\begin{equation}
\gamma^{\mu_1\ldots\mu_k}=\gamma^{\mu_1}\ldots\gamma^{\mu_k},\quad
\hbox{for}\quad 0\leq\mu_1<\cdots<\mu_k\leq3.
\label{50}
\end{equation}
Then the 16 matrices
\begin{equation}
\one,\gamma^0,\gamma^1,\gamma^2,\gamma^3,\gamma^{01},\gamma^{02},
\gamma^{03},\gamma^{12},\gamma^{13},\gamma^{23},\gamma^{012},
\gamma^{013},\gamma^{023},\gamma^{123},\gamma^{0123}
\label{60}
\end{equation}
form a basis of the matrix algebra $\M(4,\C)$
($\C$ is the field of complex numbers and $\R$ is the field of real
numbers).

\remark. We assume that the matrix $\|\eta^{\mu\nu}\|$ is independent of
$g$ from (\ref{10}). And more than this, the values $g^{\mu\nu}$ and
$\eta^{\mu\nu}$ have different mathematical meaning. Namely metric
tensor $g^{\mu\nu}$ is a geometrical object which is attribute of
Minkowski space. But the matrix $\|\eta^{\mu\nu}\|$ is an algebraic
object which contains structure constants of an underlying algebra (as
we see in a moment this algebra is the Clifford algebra). Accidentally,
the matrices $\|\eta^{\mu\nu}\|$ and $\|g^{\mu\nu}\|$ have the same
diagonal form $\diag(1,-1,-1,-1)$.


\subsection{The Clifford algebra and the spinor group.}
Let $\L$ be a real 16-dimensional vector space with
basis elements enumerated by ordered multi-indices
\begin{equation}
\ell,\ell^0,\ell^1,\ell^2,\ell^3,\ell^{01},\ell^{02},
\ell^{03},\ell^{12},\ell^{13},\ell^{23},\ell^{012},
\ell^{013},\ell^{023},\ell^{123},\ell^{0123}.
\label{80}
\end{equation}
Suppose the multiplication of elements of $\L$ is defined by the
following rules:
\begin{description}
\item[(i)] $\L$ is an associative algebra (with the unity element $\ell$)
w.r.t. this multiplication;

\item[(ii)] $\ell^\mu\ell^\nu+\ell^\nu\ell^\mu=2\eta^{\mu\nu}\ell$;

\item[(iii)] $\ell^{\mu_1}\ldots\ell^{\mu_k}=\ell^{\mu_1\ldots\mu_k},\quad
\hbox{for}\quad 0\leq\mu_1<\cdots<\mu_k\leq3$.
\end{description}
Then this algebra $\L$ is called the (real) {\sl Clifford
algebra}\footnote{The Clifford algebra was invented in 1878 by
the English mathematician W.K.Clifford \cite{Clifford}} and is denoted by
$\cl(1,3)$, where the numbers $1$ and $3$ determine the signature of the
matrix $\|\eta^{\mu\nu}\|$. The complex Clifford algebra is denoted
by $\cl_\C(1,3)$.

Elements of $\cl(1,3)$ of the form
\begin{equation}
U=\sum_{\mu_1<\cdots<\mu_k} u_{\mu_1\ldots\mu_k}\ell^{\mu_1\ldots\mu_k}
\label{90}
\end{equation}
are said to be elements of {\sl rank} $k$. For every $k=0,1,2,3,4$
the set of elements of rank $k$ is a subspace $\cl^k(1,3)$ of
$\cl(1,3)$ and
$$
\cl(1,3)=\cl^0(1,3)\oplus\ldots\oplus\cl^4(1,3)=
\cl^{\rm even}(1,3)\oplus\cl^{\rm odd}(1,3),
$$
where
\begin{eqnarray*}
\cl^{\rm even}(1,3)&=&\cl^0(1,3)\oplus\cl^2(1,3)\oplus\cl^4(1,3),\\
\cl^{\rm odd}(1,3)&=&\cl^1(1,3)\oplus\cl^3(1,3).
\end{eqnarray*}
The dimensions of the spaces $\cl^k(1,3),\, k=0,1,2,3,4$ are equal to
$1,4,6,4,1$ respectively and the dimensions of $\cl^{\rm even}(1,3)$ and
$\cl^{\rm odd}(1,3)$ are equal to $8$.
Four elements of $\cl(1,3)$ are called {\sl generators} of the Clifford
algebra if any element of $\cl(1,3)$ can be represented as a linear
combination of products of these generators.
Four generators $L^\mu\in\cl(1,3)$ are said to be {\sl primary
generators} if $L^\mu L^\nu+L^\nu L^\mu=2 \eta^{\mu\nu}\ell$.

Hence, the basis elements $\ell^\mu$ are primary generators of the
Clifford algebra $\cl(1,3)$.

Let us define the trace of a Clifford algebra element as a linear operation
$\tr\,:\,\cl\to\R$ or $\tr\,:\,\cl_\C\to\C$ such that $\tr(\ell)=1$ and
$\tr(\ell^{\mu_1\ldots\mu_k})=0,\,k=1,2,3,4$. The reader can easily prove
that
$$
\tr(UV-VU)=0,\quad \tr(V^{-1}UV)=\tr U,\quad U,V\in\cl_\C.
$$

For
$$
U=\sum_{\mu_1<\cdots<\mu_k}u_{\mu_1\ldots\mu_k}\ell^{\mu_1\ldots\mu_k}
\in\cl^k_\C(1,3)
$$
we may define an involution $*:\cl^k_\C\to\cl^k_\C$, $k=0,\ldots,4$ by
\begin{equation}
U^*=\sum_{\mu_1<\cdots<\mu_k}\bar{u}_{\mu_1\ldots\mu_k}
\ell^{\mu_k}\ldots\ell^{\mu_1},
\label{95}
\end{equation}
where the bar means complex conjugation. For $U\in\cl^k(1,3)$ we have
$$
U^*=(-1)^{\frac{k(k-1)}{2}}U.
$$
It is readily seen that
\begin{equation}
U^{**}=U,\quad (UV)^*=V^* U^*,\quad U,V\in\cl(1,3).
\label{100}
\end{equation}
Let us define the group with respect to multiplication
$$
\Spin(1,3)=\{S\in\cl^{\rm even}(1,3)\,:\,S^*S=\ell\},
$$
which is called {\sl the spinor group}.
For any $F\in\cl^{\rm even}(1,3)$ consider the linear operator
$G_F\,:\,\cl(1,3)\to\cl(1,3)$ such that
\begin{equation}
G_F(U)=F^*UF.
\label{110}
\end{equation}
In the sequel we use the following well known propositions (see proofs
in \cite{Lounesto}).

\begin{description}
\item[Proposition 1.] If $T\in\cl^{\rm even}(1,3)$ or
$T\in\cl^{\rm odd}(1,3)$, then
\begin{eqnarray*}
G_T\,&:&\,\cl^k(1,3)\to\cl^k(1,3),\quad k=1,2,3;\\
G_T\,&:&\,\cl^0(1,3)\oplus\cl^4(1,3)\to\cl^0(1,3)\oplus\cl^4(1,3).
\end{eqnarray*}
\item[Proposition 2.] If $S\in\Spin(1,3)$, then
$$
G_S\,:\,\cl^k(1,3)\to\cl^k(1,3),\quad k=0,1,2,3,4.
$$
and
\begin{equation}
S^*\ell^\nu S=p^\nu_\mu\ell^\mu,\quad\hbox{for}\quad S\in\Spin(1,3),
\label{120}
\end{equation}
where the matrix $P=\|p^\nu_\mu\|$ satisfies
the relations
\begin{equation}
P^T g P=g,\quad \det P=1, \quad p^0_0>0.
\label{130}
\end{equation}
\end{description}
\par

Therefore if we transform the coordinate system
\begin{equation}
\tilde{x}^\nu=p^\nu_\mu x^\mu,
\label{140}
\end{equation}
with the aid of this matrix $P=P(S)$, then we get the proper orthochronous
Lorentz transformation from the group \sop.
Conversely, if some matrix $P$ specifies a transformation (\ref{140})
from the group \sop, then there exist two elements $\pm S\in\Spin(1,3)$
such that the formula (\ref{120}) is satisfied (in other words
$\Spin(1,3)$ is a double
covering of \sop).
We say that the transformation of coordinates (\ref{140}),(\ref{130})
from the group \sop  is {\sl associated} with the element
$S\in\Spin(1,3)$ if (\ref{120}) holds.


\subsection{Secondary generators of the Clifford algebra.}
Denote $\ell^5=\ell^{0123}=\ell^0 \ell^1 \ell^2 \ell^3$. Then
$(\ell^5)^2=-\ell$ and $\ell^5$ commutes with all even elements and
anticommutes with all odd elements of $\cl(1,3)$.

\df 2. If elements $H\in\cl^1(1,3)$ and $I,K\in\cl^2(1,3)$ satisfy the
relations
\be
H^2=\ell,\quad I^2=K^2=-\ell,\quad [H,I]=[H,K]=0,\quad \{I,K\}=IK+KI=0,
\label{HIK}
\ee
then the elements $H,\ell^5,I,K$ are said to be {\sl secondary
generators} of
the Clifford algebra $\cl(1,3)$. \par

In particular, if we take
\be
\acute{H}=\ell^0,\quad\acute{I}=-\ell^{12},\quad\acute{K}=-\ell^{13},
\label{HIK:particular}
\ee
then these elements satisfy (\ref{HIK}) and hence, the elements
$\acute{H},\ell^5,\acute{I},\acute{K}$ are secondary generators of
$\cl(1,3)$.

If $H,\ell^5,I,K$ are secondary generators of $\cl(1,3)$, then the
16 elements
$$
\ell,H,I,K,HI,HK,IK,HIK,\ell^5,\ell^5H,\ell^5I,\ell^5K,\ell^5HI,\ell^5HK,
\ell^5IK,\ell^5HIK
$$
are the basis elements of $\cl(1,3)$ (linear independent) and the
trace $\tr$ of every element of this basis, except $\ell$, is equal
to zero.

Let $H,\ell^5,I,K$ be secondary generators of $\cl(1,3)$.
The first pair $H,\ell^5$ is such that
\be
H^2=\ell,\quad (\ell^5)^2=-\ell,\quad \{H,\ell^5\}=0.
\label{510}
\ee
Thus the elements $H,\ell^5$ are generators of the Clifford algebra
$\cl(1,1)$.

The second pair $I,K$ is such that
\be
I^2=K^2=-\ell,\quad \{I,K\}=0.
\label{520}
\ee
Therefore the elements $I,K$
are generators of the Clifford algebra $\cl(0,2)$
(which is isomorphic to the algebra of quaternions).
Furthermore, the elements $H,\ell^5$ are commute with the elements $I,K$
\be
[H,I]=[H,K]=[\ell^5,I]=[\ell^5,K]=0.
\label{530}
\ee
Consequently the Clifford algebra $\cl(1,3)$ is isomorphic to the direct
product
$$
\cl(1,3)\simeq\cl(1,1)\otimes\cl(0,2).
$$
This relation leads to the well known fact that $\cl(1,3)$ can be
represented by the algebra $\M(2,\H)$ of $2\!\times\!2$ matrices with
quaternion elements.

We omit proofs of the following three propositions.
\medskip

\noindent{\bf Proposition 3}. Suppose elements $H\in\cl^\odd(1,3)$ and
$I\in\cl^\even(1,3)$ satisfy the relations
\begin{equation}
H^2=\ell,\quad I^2=-\ell,\quad [H,I]=0.
\label{370}
\end{equation}
Then there exists an invertible element $T\in\cl^\even(1,3)$ such that
$$
T^{-1}HT=\ell^0,\quad T^{-1}IT=-\ell^{12}.
$$
\medskip

\noindent{\bf Proposition 4}. Suppose elements  $H\in\cl^1(1,3)$ and
$I\in\cl^2(1,3)$ satisfy the relations (\ref{370}). Then there exists
an element $S\in\Spin(1,3)$ such that
$$
S^*HS=\ell^0,\quad S^*IS=-\ell^{12}.
$$
\medskip

\noindent{\bf Proposition 5}. Suppose $H,\ell^5,I,K$ are secondary
generators of $\cl(1,3)$; then there exists a unique element $S\in\Spin(1,3)$
such that
$$
S^*HS=\ell^0,\quad S^*IS=-\ell^{12},\quad S^*KS=-\ell^{13}.
$$
\par
\medskip


\subsection{Idempotents, left ideals, and matrix representations of
$\cl(1,3)$.}

An element $\tau$ of an algebra ${\cal A}$ is said to be {\sl an
idempotent} if $\tau^2=\tau$. A subspace $\I\subseteq\A$ is called
{\sl a left ideal} of the algebra $\A$ if $au\in\I$ for all
$a\in\A,\,u\in\I$. Every idempotent $\tau\in\A$ generates the left ideal
$$
\I(\tau)=\{a\tau\,:\,a\in\A\}.
$$
Consider the idempotent
\begin{equation}
t=\frac{1}{4}(\ell+H)(\ell-i I)
\label{240}
\end{equation}
and the left ideal $\I(t)$ of $\cl_\C(1,3)$. The complex dimension of
$\I(t)$ is equal to $4$. Let us denote
\begin{equation}
t_k=F_k t,\quad k=1,2,3,4,
\label{250}
\end{equation}
where
\begin{equation}
F_1=\ell,\quad F_2=K,\quad F_3=-I\ell^5,\quad F_4=-KI\ell^5.
\label{260}
\end{equation}
Elements $t_k\in\I(t),\,k=1,2,3,4$ are linear independent and form a
basis of $\I(t)$.
It is easy to check that $t_k t_1=t_k$ and $t_k t_n=0$ for $n\neq1$.

Let us define an operation of Hermitian conjugation
\be
U^\dagger := HU^*H,\quad U\in\cl_\C(1,3)
\label{Hermitian:conj}
\ee
such that $(UV)^\dagger=V^\dagger U^\dagger$, $U^{\dagger\dagger}=U$.

Now we may introduce a scalar product of elements of the left ideal
$\I(t)$
\begin{equation}
(U,V) := 4\,\tr(UV^\dagger),\quad U,V\in\I(t).
\label{scalar:product}
\end{equation}
This scalar product converts the left ideal $\I(t)$ into the four
dimensional unitary space.

\theorem. The basis elements $t_k=t^k$, $k=1,2,3,4$ of $\I(t)$ are mutually
orthogonal w.r.t. the scalar product (\ref{scalar:product})
$$
(t_k,t^n)=\delta_k^n,\quad k,n=1,2,3,4
$$
where $\delta_k^k=1$ and $\delta_k^n=0$ for $k\neq n$.

\proof\,\, is by direct calculation.
\medskip

In what follows we use the formulas
\be
\Psi=(\Psi,t^k)t_k\quad\hbox{for}\quad\Psi\in\I(t)
\label{ideal:decomposition}
\ee
and
$$(KU,V)=(U,K^\dagger V)\quad\hbox{for}\quad
U,V\in\I(t),\,K\in\cl_\C(1,3).
$$

Now we introduce the following concept. We claim that the set of
secondary generators $H,\ell^5,I,K$ uniquely defines a matrix representation
for $\cl(1,3)$.
Indeed, for $U\in\cl(1,3)$ the products
$U t_k$ belong to $\I(t)$ and can be represented as linear
combinations of basis elements $t_n$ with certain coefficients
$\gamma(U)^n_k$
\begin{equation}
Ut_k=\gamma(U)^n_k t_n,\quad k=1,2,3,4.
\label{270}
\end{equation}
Thus, the matrix $\gamma(U)$ with the elements $\gamma(U)^n_k$
(an upper index enumerates lines and a lower index enumerate columns of
a matrix)
is associated with the element $U\in\cl(1,3)$. In particular, the matrices
$\gamma^\mu=\gamma(\ell^\mu)$ are defined by the formulas
\be
\ell^\mu t_k=\gamma(\ell^\mu)^n_k t_n,\quad k=1,\ldots,4;\,\,
\mu=0,\ldots,3.
\label{272}
\ee
Considering scalar products of the left and right hand sides of
(\ref{270}),(\ref{272}) by $t^n$
and using mutual orthogonality of $t_k$, we get
\be
\gamma(U)^n_k=(Ut_k,t^n)
\label{275}
\ee
and, in particular
\be
\gamma^{n\mu}_k=\gamma(\ell^\mu)^n_k=(\ell^\mu t_k,t^n).
\label{277}
\ee
It can be shown that
$$
\gamma(UV)=\gamma(U)\gamma(V),\quad
\gamma(\ell)=\one,\quad \gamma(\alpha U)=\alpha\gamma(U),\,\alpha\in\C.
$$
Therefore the map $\gamma\,:\,\cl\to\M(4,\C)$ defined by (\ref{270}) is a
matrix representation of the Clifford algebra $\cl$.

If we take secondary generators of the form
(\ref{HIK:particular}), then the matrices $\gamma^\mu$ from
(\ref{277}) are equal
to the matrices from (\ref{30}).

Let $S$ be an element of the group $\spin(1,3)$ and $\gamma(S)$,
$\gamma(S^*)$ be the matrix representation of $S$, $S^*$ given by
(\ref{270}). From the formula  (\ref{120}) we have
\be
S^*\ell^\mu S=p^\mu_\nu \ell^\nu,\quad
S\ell^\nu S^*=q^\nu_\mu \ell^\mu,
\label{pq}
\ee
where
$$
p^\mu_\nu q^\nu_\lambda=\delta^\mu_\lambda,\quad
q^\mu_\nu p^\nu_\lambda=\delta^\mu_\lambda.
$$
For the secondary generators $H,\ell^5,I,K$ consider the transformation
$$
(H,\ell^5,I,K)\to(S^*HS,\ell^5,S^*IS,S^*KS),
$$
which leads to the transformation of the left ideal
$\I(t)\to\I(S^*tS)$ and the basis elements
$$
t_k\to \acute{t_k}=S^*t_k S.
$$
Now we may define a new matrix representation of the Clifford algebra
$\acute{\gamma}\,:\,\cl\to\M(4,\C)$ with the aid of the formula
\be
U\acute{t}_k=\acute{\gamma}(U)^n_k \acute{t}_n.
\label{prime:representation}
\ee

\theorem. For every $U\in\cl$ the matrix $\gamma(U)$ defined with the
aid of (\ref{275}) connected with the matrix representation
$\acute{\gamma}(U)$ by the formula
$$
\acute{\gamma}(U)=\gamma(S)\gamma(U)\gamma(S^*).
$$
\par

\proof. It is sufficient to prove this theorem for the primary
generators $\ell^\mu$. We have
$$
\ell^\mu t_k=\gamma(\ell^\mu)^n_k t_n.
$$
Multiplying both sides of this relation from the left by $S^*$ and from
the right by $S$, we obtain
$$
(S^*\ell^\mu S) (S^*t_k S)=\gamma(\ell^\mu)^n_k (S^*t_n S).
$$
Substituting $S^*t_k S=\acute{t}_k$ and $S^*\ell^\mu
S=p^\mu_\nu\ell^\nu$ from (\ref{pq}), we get
$$
p^\mu_\nu\ell^\nu\acute{t}_k=\gamma(\ell^\mu)^n_k\acute{t}_n.
$$
Multiplying both sides by $q^\lambda_\mu$ from (\ref{pq}) and summing
over $\mu$, we obtain
\begin{eqnarray*}
q^\lambda_\mu p^\mu_\nu \ell^\nu\acute{t}_k=\ell^\lambda\acute{t}_k&=&\\
&=&\gamma(q^\lambda_\mu\ell^\mu)^n_k\acute{t}_n=\gamma(S\ell^\lambda
S^*)^n_k\acute{t}_n\\
&=&(\gamma(S)\gamma(\ell^\lambda)\gamma(S^*))^n_k
\acute{t}_n.
\end{eqnarray*}
This completes the proof.
\bigskip

Let us note that $\acute{\gamma}(S)=\gamma(S)$ and
 $\acute{\gamma}(S^*)=\gamma(S^*)$


\subsection{A one-to-one correspondence between $\I(t)$ and
$\cl_\C(1,3)$.}
The dimension of the linear space $\cl^\even(1,3)$  is equal to $8$. The
left ideal $\I(t)$ has the complex dimension $4$ and, thus, the real
dimension $8$. We may consider the map $\cl^\even(1,3)\to\I(t)$ given by
the formula
$$
\Psi\to\Psi t=\phi^k t_k,
$$
where $\Psi\in\cl^\even(1,3)$ and $\phi^k=(\Psi t,t^k)$. Let us prove
that this map gives the one-to-one correspondence between the even
Clifford algebra $\cl^\even(1,3)$ and the left ideal $\I(t)$.

\theorem 3. Suppose $\Phi=\phi^k t_k\in\I(t)$. Then the
equation for $\Omega\in\cl^\even(1,3)$
\begin{equation}
\Omega t=\Phi
\label{330}
\end{equation}
has a unique solution
\begin{equation}
\Omega=F_k(\alpha^k\ell+\beta^k I),
\label{340}
\end{equation}
where $\phi^k=\alpha^k+i\beta^k$ and $F_k$ are defined in (\ref{260}).
\medskip

\proof. Multiplying both sides of (\ref{340}) from the right by $t$ and
using the relation $I t=it$, we see that the formula (\ref{340})
really gives the solution of (\ref{330}).
Now we must prove that the homogeneous equation $\Omega t=0$ has only
trivial solution $\Omega=0$.
Firstly we give the proof for the secondary generators from
(\ref{HIK:particular}).
Suppose $\Omega$ is of the form
$$
\Omega=\omega\ell+\sum_{\mu<\nu}\omega_{\mu\nu}\ell^{\mu\nu}+
\omega_5\ell^5.
$$
Then the element $\Omega t\in\I(t)$ can be expanded in the basis $t_k$
$$
\Omega t=(\omega-i\omega_{12})t_1+(-\omega_{13}-i\omega_{23})t_2+
(-\omega_{03}+i\omega_5)t_3+(-\omega_{01}-i\omega_{02})t_4.
$$
Therefore from $\Omega t=0$ we get $\Omega=0$.
For the case of general secondary generators we must represent
$\Omega$ as the linear combination of products of the generators
$\ell^5,I,K$ and  arguing as above.
So the solution
(\ref{340}) of (\ref{330}) is unique. This completes the proof.
\medskip

\theorem. If $\Psi\in\cl^\even(1,3)$ and $S\in\Spin(1,3)$, then the
relation
\be
\Psi t=\psi^k t_k,
\label{delta}
\ee
where $\psi^k=(\Psi t,t^k)\in\C$, is invariant under the following
transformation:
\begin{eqnarray}
t_k &\to& \acute{t}_k=S^*t_k S,\nonumber\\
\Psi &\to& \acute{\Psi}=\Psi S,\label{Spin:transformation}\\
\psi^k &\to& \acute{\psi}^k=\gamma(S)^k_l
\psi^l=(St_l,t^k)\psi^l.\nonumber
\end{eqnarray}

\proof.\,\, We have
\begin{eqnarray*}
(\acute{\Psi}\acute{t}-\acute{\psi}^k \acute{t}_k)S^* &=&
\Psi t-(St_l,t^k)\psi^l S^* t_k=\\
&=& \Psi t-\psi^l(St_l,t^k)(S^*t_k,t^n)t_n=\\
&=& \Psi t-\psi^n t_n.
\end{eqnarray*}
Here we use the formulas
$$
S^*t_k=(S^*t_k,t^n)t_n
$$
and
$$
(St_l,t^k)(S^*t_k,t^n)=\gamma(S)^k_l \gamma(S^*)^n_k=\delta^n_l.
$$
This completes the proof.
\bigskip


Consider the correspondence $\ell^\mu\to\gamma^\mu$ between primary
generators of the Clifford algebra and Dirac's $\gamma$-matrices. Let us
extend this correspondence to the map $\gamma\,:\,\cl(1,3)\to\M(4,\C)$
or $\gamma\,:\,\cl_\C(1,3)\to\M(4,\C)$ in such a way that any element
of the Clifford algebra
\begin{equation}
U=u\ell+u_\mu\ell^\mu+\sum_{\mu_1<\mu_2}u_{\mu_1\mu_2}\ell^{\mu_1\mu_2}+
\sum_{\mu_1<\mu_2<\mu_3}u_{\mu_1\mu_2\mu_3}\ell^{\mu_1\mu_2\mu_3}+
u_{0123}\ell^{0123}
\label{150}
\end{equation}
corresponds to the matrix ${\bf U}\in\M(4,\C)$
\begin{equation}
{\bf U}=u\one+u_\mu\gamma^\mu+
\sum_{\mu_1<\mu_2}u_{\mu_1\mu_2}\gamma^{\mu_1\mu_2}+
\sum_{\mu_1<\mu_2<\mu_3}u_{\mu_1\mu_2\mu_3}\gamma^{\mu_1\mu_2\mu_3}+
u_{0123}\gamma^{0123}
\label{160}
\end{equation}
And let $\ell\,:\,\M(4,\C)\to\cl_\C(1,3)$ or
$\ell\,:\,\M(4,\C)\to\cl(1,3)$ be the inverse map such that any matrix
(\ref{160}) corresponds to the element (\ref{150}) of the (complex)
Clifford algebra.


\subsection{The covariance of the Dirac equation.}
Let us consider the transformation of the Dirac equation (\ref{20}) under
a change of coordinates $(x)\to(\tilde{x})$ from the group \sop. This
change of coordinates (\ref{140}),(\ref{130}) is associated with an
element $S\in\Spin(1,3)$ in accordance with (\ref{120}). By
${\bf S}=\gamma(S)$ denote the matrix representation of the element $S$.
Then the matrix ${\bf S}$ satisfies
\begin{equation}
{\bf S}^*\gamma^\nu{\bf S}=p^\nu_\mu\gamma^\mu.
\label{170}
\end{equation}
The covectors $\partial_\mu$ and $a_\mu$ are transformed under the change of
coordinates (\ref{140}) as
\begin{equation}
\partial_\mu=p^\nu_\mu\tilde{\partial}_\nu,\quad
a_\mu=p^\nu_\mu\tilde{a}_\nu,
\label{180}
\end{equation}
where $\tilde{\partial}_\nu=\partial/\partial\tilde{x}^\nu$ and
$\tilde{a}_\nu$ are components of the covector $a_\mu$ in coordinates
$(\tilde{x})$. By substituting (\ref{180}),(\ref{170}) into (\ref{20}),
we obtain
\begin{eqnarray*}
\gamma^\mu(\partial_\mu \psi+i a_\mu \psi)+i m \psi
&=&p^\nu_\mu\gamma^\mu(\tilde{\partial}_\nu\psi+
i\tilde{a}_\nu\psi)+i m\psi=\\
&=&{\bf S}^*\gamma^\nu{\bf S}(\tilde{\partial}_\nu\psi+
i\tilde{a}_\nu\psi)+i m\psi=\\
&=&{\bf S}^*(\gamma^\nu(\tilde{\partial}_\nu({\bf S}\psi)+
i\tilde{a}_\nu({\bf S}\psi))+i m({\bf S}\psi))
\end{eqnarray*}

Hence, if a column of four complex valued functions $\psi=\psi(x)$ in the
coordinates $(x)$ satisfies the equation (\ref{20}), then the column
$\tilde{\psi}={\bf S}\psi(x(\tilde{x}))$ in coordinates $(\tilde{x})$
satisfies the equation
\begin{equation}
\gamma^\nu(\tilde{\partial}_\nu\tilde{\psi}+i\tilde{a}_\nu\tilde{\psi})+
i m\tilde{\psi}=0,
\label{190}
\end{equation}
which has the same form as (\ref{20}).

\df 1. A column of four complex valued functions $\psi$ is called
{\sl a bispinor} if $\psi$ transforms under the change of
coordinates (\ref{140}),(\ref{130}),(\ref{120}) as
$\psi\to\tilde{\psi}={\bf S}\psi(x(\tilde{x}))$, where ${\bf
S}=\gamma(S)$.\par


\subsection{Algebraic bispinors and the Dirac equation.}
Consider the equation \cite{Reisz}
\begin{equation}
\ell^\mu(\partial_\mu\rho+i a_\mu\rho)+im\rho=0,
\label{290}
\end{equation}
where $\rho=\rho(x)\in\cl_\C(1,3)$. The element $\rho$ has 16 complex
components, i.e., four times more than the bispinor. Multiplying the
equation (\ref{290}) from the right by $t$, we obtain that the element
$\theta=\rho t\in\I(t)$ satisfying the same equation
\begin{equation}
\ell^\mu(\partial_\mu\theta+i a_\mu\theta)+im\theta=0.
\label{300}
\end{equation}

\theorem 2. An element $\theta=\psi_k t^k\in\I(t)$ satisfies the equation
(\ref{300}) iff the column $\psi=(\psi_1\,\psi_2\,\psi_3\,\psi_4)^T$
satisfies the Dirac equation (\ref{20}).\par

\proof. Necessity. Substituting $\theta=\psi_k t^k$ into (\ref{300}) and
using (\ref{270}), we get
\begin{equation}
\ell^\mu t^k(\partial_\mu\psi_k+i a_\mu\psi_k)+im\psi_l
t^l=(\gamma^{\mu k}_l(\partial_\mu\psi_k+
ia_\mu\psi_k)+im\psi_l)t^l=0.
\label{310}
\end{equation}
Taking into account the linear independence of $t^k$, we obtain the
equations
$$
\gamma^{\mu k}_l(\partial_\mu\psi_k+
ia_\mu\psi_k)+im\psi_l=0,\quad l=1,2,3,4,
$$
which is evidently equivalent to the Dirac equation (\ref{20}).

Arguing as above but in inverse order, we prove sufficiency.
This completes the proof.
\medskip

We say that for $t$ from (\ref{240})
the left ideal $\I(t)\subset\cl_\C(1,3)$ is
{\sl the spinor space}.
Elements of the spinor space are called algebraic bispinors.
The formula $\theta=\psi_k t^k$ gives the
relation between the algebraic bispinor $\theta\in\I(t)$ and the bispinor
$\psi=(\psi_1\,\psi_2\,\psi_3\,\psi_4)^T$.


\subsection{Hestenes' form of the Dirac equation.}
Let $H,\ell^5,I,K$ be secondary generators of $\cl(1,3)$ and be independent
of $x$.
Consider the equation  for $\Psi=\Psi(x)\in\cl^\even(1,3)$
\begin{equation}
\ell^\mu(\partial_\mu\Psi+a_\mu\Psi I)+m\Psi H I=0,
\label{320}
\end{equation}
which was invented by D.Hestenes \cite{Hestenes} in 1966. This
equation is called {\sl Hestenes' form of the Dirac equation} (HDE).
Let us show that
the equation (\ref{320}) is equivalent to the equation (\ref{300}) and
consequently to the Dirac equation (\ref{20}). To prove this fact we
need the following theorem.

\theorem 4. An element $\Psi=\Psi(x)\in\cl^\even(1,3)$ satisfies the
equation (\ref{320}) iff the element $\theta=\Psi t\in\I(t)$ satisfies the
equation (\ref{300}). \par

\proof. Necessity. Suppose $\Psi\in\cl^\even(1,3)$ satisfies the
equation (\ref{320}). Let us multiply (\ref{320}) from the right by the
idempotent $t$ and use the relations
$$
H t=t,\quad I t=it.
$$
Then we get the equation (\ref{300}) for $\theta=\Psi t\in\I(t)$
$$
\ell^\mu(\partial_\mu\Psi t+a_\mu\Psi I t)+m\Psi H I t=
\ell^\mu(\partial_\mu\Psi t+i a_\mu\Psi t)+i m\Psi t=0.
$$
Sufficiency. Suppose the element $\theta\in\I(t)$ satisfies the equation
(\ref{300}). Let us multiply this equation from the right by $t$ and use the
relations
$$
t=t H,\quad it=tI,\quad \theta t=\theta.
$$
Then we obtain
\begin{equation}
\ell^\mu(\partial_\mu\theta+a_\mu\theta I)+m\theta H I=0.
\label{350}
\end{equation}
Now using Theorem 3 we may take $\Psi\in\cl^\even(1,3)$ as the solution
of the equation $\Psi t=\theta$. We claim that this $\Psi$ satisfies the
equation (\ref{320}). In fact, substituting $\theta=\Psi t$ into the
equation (\ref{350}) and multiplying from the left by $H$, we get
$$
0=H(\ell^\mu(\partial_\mu\Psi+a_\mu\Psi I)+m\Psi H I)t\equiv
\Omega t,
$$
where $\Omega\in\cl^\even(1,3)$. By Theorem 3 the equation $\Omega
t=0$ has only the trivial solution $\Omega=0$. Hence $\Psi$ satisfies
(\ref{320}). The theorem is proved.
\medskip

Thus, we have proved that HDE (\ref{320}) for
$\Psi\in\cl^\even(1,3)$ is equivalent to the Dirac equation (\ref{20})
for $\psi=(\psi_1\,\psi_2\,\psi_3\,\psi_4)^T\,$,
$\psi_k=\alpha_k+i\beta_k$ and the relation between
these solutions is
\begin{equation}
\Psi=F^k(\alpha_k\ell+\beta_k I),
\label{360}
\end{equation}
where the summation over $k=1,2,3,4$ is assumed.

Consider a change of coordinates (\ref{140}),(\ref{130}),(\ref{120})
from the group \sop associated with the element $S\in\Spin(1,3)$.
Arguing as in section 3, we see that HDE
transforms under this change of coordinates as follows
$$
\ell^\mu(\partial_\mu\Psi+a_\mu\Psi I)+m\Psi H I=
S^*(\ell^\nu(\tilde{\partial}_\nu(S\Psi)+\tilde{a}_\nu(S\Psi)I)+
m(S\Psi)H I).
$$
Hence, if $\Psi=\Psi(x)\in\cl^\even(1,3)$ in coordinates $(x)$ satisfies
HDE (\ref{320}), then  the element
$\tilde{\Psi}=S\Psi(x(\tilde{x}))$ in coordinates $(\tilde{x})$ satisfies
the equation
$$
\ell^\nu(\tilde{\partial}_\nu\tilde{\Psi}+\tilde{a}_\nu\tilde{\Psi}I)+
m\tilde{\Psi}H I=0,
$$
which has the same form as (\ref{320}). As was shown, the relation in
coordinates $(x)$ between a solution $\psi$ of the Dirac equation and
the solution $\Psi$ of HDE is given by the
formula
$$
\Psi t =\psi_k t^k,\quad \psi=(\psi_1\,\psi_2\,\psi_3\,\psi_4)^T.
$$
And the relation in coordinates $(\tilde{x})$ between the solution
$\tilde{\psi}={\bf S}\psi$ of the Dirac equation and the solution
$\tilde{\Psi}=S\Psi$ of HDE is given by the
formula
$$
\tilde{\Psi} t =\tilde{\psi}_l t^l,\quad \tilde{\psi}=
(\tilde{\psi}_1\,\tilde{\psi}_2\,\tilde{\psi}_3\,\tilde{\psi}_4)^T.
$$
Indeed, using the formula $St=\gamma(S)^k_l t^l$, where $\gamma(S)^k_l$
are elements of the matrix ${\bf S}$, we obtain
$$
\tilde{\Psi}t=S\Psi t=\psi_k S t^k=\psi_k\gamma(S)^k_l
t^l=\tilde{\psi}_l t^l.
$$
\medskip

\subsection{The Grassmann-Clifford bialgebra.}
Suppose that for elements of $\cl(1,3)$ the exterior
multiplication (denoted by $\wedge$) is defined by the following rules:
\begin{description}
\item[(j)] $\cl(1,3)$ is an associative algebra (with the unity
element $\ell$) with respect to exterior multiplication;
\item[(jj)] $\ell^\mu\wedge\ell^\nu=-\ell^\nu\wedge\ell^\mu\,\,$,
$\mu,\nu=0,1,2,3$;
\item[(jjj)] $\ell^{\mu_1}\ww\ell^{\mu_k}=\ell^{\mu_1\ldots\mu_k}$ for
$0\leq\mu_1<\cdots<\mu_k\leq3$.
\end{description}
The resulting algebra (equipped with the Clifford multiplication and with
the exterior multiplication) is called {\sl the Grassmann-Clifford
bialgebra} and is denoted by $\Lambda(1,3)$. The complex valued
Grassmann-Clifford bialgebra is denoted by  $\Lambda_\C(1,3)$. Any
element $U\in\Lambda(1,3)$ can be expanded in the basis as in
(\ref{150}).
The coefficients $u_{\mu_1\ldots\mu_k}$ in (\ref{150}) are
enumerated by ordered
multi-indices. Let us take the coefficients that are
antisymmetric w.r.t. all indices
$$
u_{\mu_1\ldots\mu_k}=u_{[\mu_1\ldots\mu_k]},
$$
where square brackets denote the operation of alternation (with the
division by $k!$). Then elements of the form
\begin{equation}
\sum_{\mu_1<\cdots<\mu_k}u_{\mu_1\ldots\mu_k}\ell^{\mu_1\ldots\mu_k}=
\frac{1}{k!}u_{\nu_1\ldots\nu_k}\ell^{\nu_1}\ww\ell^{\nu_k}=
\frac{1}{k!}u_{\nu_1\ldots\nu_k}\ell^{\nu_1}\ldots\ell^{\nu_k}
\label{530}
\end{equation}
are said to be elements of rank $k$ and belong to $\Lambda^k(1,3)$,
where $\Lambda^k(1,3)$, $\Lambda^\even(1,3)$, $\Lambda^\odd(1,3)$ are
the same as $\cl^k(1,3)$, $\cl^\even(1,3)$, $\cl^\odd(1,3)$.
If $U\in\Lambda^r(1,3),\,V\in\Lambda^s(1,3)$ then
\begin{equation}
U\wedge V=(-1)^{rs}V\wedge U\in\Lambda^{r+s}(1,3).
\label{540}
\end{equation}


\section{Part II.}
\subsection{The exterior algebra of Minkowski space.}
Let $\E$ be Minkowski space with the metric tensor (\ref{10}), with
coordinates $x^\mu$, with basis coordinate vectors $e_\mu$, and with
basis covectors $e^\mu=g^{\mu\nu}e_\nu$. Consider a covariant
antisymmetric tensor field of rank $0\leq k\leq3$ on $\E$ with
components
$$
u_{\mu_1\ldots\mu_k}=u_{[\mu_1\ldots\mu_k]}(x).
$$
It is suitable to write this field with the aid of the expression
\begin{equation}
\frac{1}{k!}u_{\nu_1\ldots\nu_k}e^{\nu_1}\ww e^{\nu_k},
\label{560}
\end{equation}
where the expression $e^{\nu_1}\ww e^{\nu_k}$ in the fixed coordinates
$(x)$ can be considered as an element of the Grassmann algebra.
Under a linear change of coordinates
\begin{equation}
x^\mu=\frac{\partial x^\mu}{\partial\tilde{x}^\nu}\tilde{x}^\nu=
q^\mu_\nu\tilde{x}^\nu
\label{565}
\end{equation}
the transformation rule for the expression $e^{\nu_1}\ww e^{\nu_k}$
corresponds to the transformation rule for basis covectors $e^\nu$.
That is,
\begin{equation}
e^\nu=q^\nu_\mu\tilde{e}^\mu,\quad e^{\nu_1}\ww e^{\nu_k}=
q^{\nu_1}_{\mu_1}\ldots
q^{\nu_k}_{\mu_k}\tilde{e}^{\mu_1}\ww\tilde{e}^{\mu_k}.
\label{570}
\end{equation}
Therefore under the change of coordinates (\ref{565}) the expression
(\ref{560}) is invariant
$$
\frac{1}{k!}u_{\nu_1\ldots\nu_k}e^{\nu_1}\ww e^{\nu_k}=
\frac{1}{k!}\tilde{u}_{\mu_1\ldots\mu_k}\tilde{e}^{\mu_1}\ww
\tilde{e}^{\mu_k},
$$
where
$$
\tilde{u}_{\mu_1\ldots\mu_k}=p^{\lambda_1}_{\nu_1}\ldots
p^{\lambda_k}_{\nu_k} u_{\lambda_1\ldots\lambda_k}
$$
are components of the tensor field $u_{\nu_1\ldots\nu_k}$ in
the coordinates $(\tilde{x})$ and
$$
p^\lambda_\nu=\frac{\partial\tilde{x}^\lambda}{\partial x^\nu},\quad
p^\lambda_\nu q^\nu_\mu=\delta^\lambda_\mu\quad
(\delta^\lambda_\lambda=1,\,
\delta^\lambda_\mu=0\,\,\hbox{for}\,\,\mu\neq\lambda).
$$
The expressions (\ref{560}) are called  {\sl exterior forms
of rank $k$} or {\sl $k$-forms}. The set of all $k$-forms is denoted by
$\Lambda^k(\E)$. The  formal sum of $k$-forms
\begin{equation}
\sum^4_{k=0}\frac{1}{k!}u_{\nu_1\ldots\nu_k}e^{\nu_1}\ww e^{\nu_k}
\label{590}
\end{equation}
are said to be (nonhomogeneous) exterior form. The set of all exterior forms is denoted by
$\Lambda(\E)$ and
$$
\Lambda(\E)=\Lambda^0(\E)\oplus\ldots\oplus\Lambda^4(\E)=
\Lambda^\even(\E)\oplus\Lambda^\odd(\E).
$$
It is well known that the
exterior product of exterior forms is an exterior form.

Consider the Hodge star operator
$\star\,:\,\Lambda^k(\E)\to\Lambda^{4-k}(\E)$. If $U\in\Lambda^k(\E)$
has the form (\ref{560}), then
$$
\star U=
\frac{1}{k!(4-k)!}\varepsilon_{\mu_1\ldots\mu_4}u^{\mu_1\ldots\mu_k}
e^{\mu_{k+1}}\ww e^{\mu_4},
$$
where
$$
u^{\mu_1\ldots\mu_k}=g^{\mu_1\nu_1}\ldots g^{\mu_k\nu_k}u_{\nu_1\ldots\nu_k}
$$
and $\varepsilon_{\mu_1\ldots\mu_4}$ is the sign of the permutation
$(\mu_1\ldots\mu_4)$.
$\star U$ is an exterior form (a covariant antisymmetric tensor) w.r.t. any change
of coordinates with a positive Jacobian.
\medskip

\remark. In this paper we consider changes of coordinates only from the
group \sop (a Jacobian is equal to 1) and do not distinguish tensors and
pseudotensors.
\medskip

It is easy to prove that for any $U\in\Lambda^k(\E)$
\begin{equation}
\star(\star U)=(-1)^{k+1}U.
\label{600}
\end{equation}
Further on we consider the bilinear operator
$\com\,:\,\Lambda^2(\E)\times\Lambda^2(\E)\to\Lambda^2(\E)$ such that
for basis 2-forms
$$
\com(e^{\mu_1}\wedge e^{\mu_2},e^{\nu_1}\wedge e^{\nu_2})=
-2g^{\mu_1 \nu_1}e^{\mu_2}\wedge e^{\nu_2}-2g^{\mu_2 \nu_2}e^{\mu_1}\wedge e^{\nu_1}
+2g^{\mu_1 \nu_2}e^{\mu_2}\wedge e^{\nu_1}+2g^{\mu_2 \nu_1}e^{\mu_1}\wedge e^{\nu_2}
$$
Evidently
$$
\com(U,V)=-\com(V,U),\quad U,V\in\Lambda^2(\E).
$$
Now we define the Clifford multiplication of exterior forms with the aid of
the following formulas:
\begin{eqnarray*}
\s{0}{U}\s{k}{V}&=&\s{k}{V}\s{0}{U}=\s{0}{U}\wedge\s{k}{V}=\s{k}{V}\wedge\s{0}{U},\\
\s{1}{U}\s{k}{V}&=&\s{1}{U} \wedge  \s{k}{V}-\star (\s{1}{U} \wedge  \star \s{k}{V}),\\
\s{k}{U}\s{1}{V}&=&\s{k}{U} \wedge  \s{1}{V}+\star (\star\s{k}{U} \wedge   \s{1}{V}),\\
\s{2}{U}\s{2}{V}&=&\s{2}{U}\wedge\s{2}{V}+
\star(\s{2}{U}\wedge\star\s{2}{V})+\frac{1}{2}\com(\s{2}{U},\s{2}{V}),\\
\s{2}{U}\s{3}{V}&=&\star \s{2}{U} \wedge  \star \s{3}{V}-\star (\s{2}{U} \wedge  \star \s{3}{V}),\\
\s{2}{U}\s{4}{V}&=&\star \s{2}{U} \wedge  \star \s{4}{V},\\
\s{3}{U}\s{2}{V}&=&-\star \s{3}{U} \wedge  \star \s{2}{V}-\star (\star \s{3}{U} \wedge  \s{2}{V}),\\
\s{3}{U}\s{3}{V}&=&\star \s{3}{U} \wedge  \star \s{3}{V}+\star (\s{3}{U} \wedge  \star \s{3}{V}),\\
\s{3}{U}\s{4}{V}&=&\star \s{3}{U} \wedge  \star \s{4}{V},\\
\s{4}{U}\s{2}{V}&=&\star \s{4}{U} \wedge  \star \s{2}{V},\\
\s{4}{U}\s{3}{V}&=&-\star \s{4}{U} \wedge  \star \s{3}{V},\\
\s{4}{U}\s{4}{V}&=&-\star \s{4}{U} \wedge  \star \s{4}{V},
\end{eqnarray*}
where ranks of exterior forms are denoted as $\s{k}{U}\in\Lambda^k(\E)$ and
$k=0,1,2,3,4$.

From this definition we may get some properties of the Clifford
multiplication of exterior forms.
\begin{description}
\item[1.] If $U,V\in\Lambda(\E)$, then $UV\in\Lambda(\E)$.
\item[2.] The axioms of associativity and distributivity are satisfied
for the Clifford multiplication.
\item[3.] $e^\mu e^\nu=e^\mu\wedge e^\nu+g^{\mu\nu}e,\quad
e^\mu e^\nu+e^\nu e^\mu=2 g^{\mu\nu}e$.
\item[4.] $e^{\mu_1}\ldots e^{\mu_k}=e^{\mu_1}\ww e^{\mu_k}
=e^{\mu_1\ldots\mu_k}\,$ for
$0\leq\mu_1<\cdots<\mu_k\leq3$.
\item[5.] If $U,V\in\Lambda^2(\E)$, then $\com(U,V)=UV-VU$.
\end{description}

Taking into account these properties of Clifford multiplication, we may
conclude that Propositions 1 -- 5 of Part I initially
formulated for elements of $\cl(1,3)$ are also valid for elements of
$\LE$.

In the sequel, we use the group
\begin{equation}
\Spin(\E)=\{S\in\Lambda^\even(\E)\,:\,S^*S=e\}.
\label{630}
\end{equation}


\subsection{Operators $d,\delta,\Upsilon,\Delta$.}
First consider the operator
$$
dV=e^\mu\wedge\partial_\mu V, \quad V\in\LE
$$
such that
\begin{description}
\item[1)] $d\,:\,\LkE\to\Lambda^{k+1}(\E)$;
\item[2)] $d^2=0$;
\item[3)] $d(U\wedge V)=dU\wedge V+(-1)^k U\wedge dV$ for
$U\in\LkE,V\in\LE$.
\end{description}

Secondly, consider the operator $\delta$
$$
\delta U=\star d\star U\quad\hbox{for}\quad U\in\LE
$$
such that
\begin{description}
\item[1)] $\delta\,:\,\LkE\to\Lambda^{k-1}(\E)$;
\item[2)] $\delta^2=0$.
\end{description}

Thirdly, consider the operator (Upsilon)
$$
\Upsilon=d-\delta
$$
such that
\begin{description}
\item[1)] $\Upsilon\,:\,\LkE\to\Lambda^{k+1}(\E)\oplus\Lambda^{k-1}(\E)$;
\item[2)] $\Upsilon U=e^\mu\partial_\mu U$.
\end{description}
The second property in this list follows from the definition of Clifford
multiplication
$$
\s{1}{U}\s{k}{V}=\s{1}{U}\wedge \s{k}{V}-\star(\s{1}{U}\wedge\star\s{k}{V})
$$
if we formally substitute $\s{1}{U}=e^\mu\partial_\mu$.

Finally, consider the Beltrami-Laplace operator
$$
\Delta=\Upsilon^2=(d-\delta)^2=-(d\delta+\delta
d)=g^{\mu\nu}\partial_\mu\partial_\nu
$$
such that
\begin{description}
\item[1)] $\Delta\,:\,\LkE\to\Lambda^k(\E)$;
\item[2)] $\Delta$ commutes with the operators $d,\delta,\Upsilon,\star$.
\end{description}


\subsection{A tensor form of the Dirac equation.}
Let $H\in\Lambda^1(\E),\,I\in\Lambda^2(\E)$ be two independent of
$x$ exterior forms such that
\begin{equation}
H^2=e,\quad I^2=-e,\quad [H,I]=0.
\label{1000}
\end{equation}
Now we consider the equation
\begin{equation}
\Upsilon\Phi+A\Phi I+m\Phi HI=0,
\label{1010}
\end{equation}
where $\Phi=\Phi(x)\in\Lambda^\even(\E)$,
$A=a_\mu(x)e^\mu\in\Lambda^1(\E)$, and $m\geq0$ is a real constant.
All the values $(\Phi,A,I,H)$ in (\ref{1010}) are exterior forms (covariant
antisymmetric tensors). Two operations are used in (\ref{1010}). Namely
the differential operator $\Upsilon=d-\delta=e^\mu\partial_\mu$ and the
Clifford multiplication of exterior forms. Both operations take exterior forms to exterior forms.
In other words, (\ref{1010}) is a tensor equation.
We say  that the equation (\ref{1010}) is {\sl the Tensor form of the Dirac
Equation} (TDE). The TDE is invariant under the following global
(independent of $x$) transformation
\begin{eqnarray}
\Phi&\to&\Phi S,\nonumber\\
H&\to&S^*HS,\label{1020}\\
I&\to&S^*IS,\nonumber
\end{eqnarray}
where $S\in\Spin(\E)$, $\partial_\mu S=0$.

In a fixed coordinate system $(x)$  the TDE is equivalent to
HDE (\ref{320}) and the connection between
(\ref{1010}) and (\ref{320}) is given by the formula
$$
(\Phi)_{e^\mu\to\ell^\mu}=\Psi.
$$
 Let us remind that  HDE
(\ref{320}) is equivalent to the Dirac equation (\ref{20}) and the
connection between them is given by the formula (\ref{360}).
\medskip

\remark. Taking into account the theorem 5, it is clear that we may use in
the TDE
$H\in\Lambda^\odd(\E),\,I\in\Lambda^\even(\E)$ which satisfies (\ref{1000})
instead of $H\in\Lambda^1(\E),\,I\in\Lambda^2(\E)$.
In this case the equation (\ref{1010}) is invariant under the global
transformation
$$
\Phi\to\Phi T,\quad H\to T^{-1}HT,\quad I\to T^{-1}IT,
$$
where $T\in\Lambda^\even(\E)$ is an invertible element.
\medskip

Under a linear change of coordinates $(x)\to(\tilde{x})$ all exterior forms
are invariants. Therefore in coordinates $(\t{x})$ the TDE has the form
$$
\t{\Upsilon}\t{\Phi}+\t{A}\t{\Phi}\t{I}+m\t{\Phi}\t{H}\t{I}=0,
$$
where $\t{\Upsilon}=\t{e}^\mu\partial/\partial\t{x}^\mu=\Upsilon$,
$\t{\Phi}=\Phi$, $\t{A}=A$, $\t{H}=H$, $\t{I}=I$ and
$\t{\Phi},\t{A},\t{H},\t{I}$ are the exterior forms written in coordinates
$(\t{x})$.

Let us define the trace of an exterior form as a linear operation
$\tr\,:\,\LE\to\R$ or $\tr\,:\,\LCE\to\C$ such that $\tr(e)=1$ and
$\tr(e^{\mu_1\ldots\mu_k})=0,\,k=1,2,3,4$. The reader can easily prove
that
$$
\tr(UV-VU)=0,\quad \tr(V^{-1}UV)=\tr U,\quad U,V\in\LE.
$$
Now we can find the conservative current for the TDE. For this let us denote
$$
C=\Phi^*(\Upsilon\Phi+A\Phi I+m\Phi HI),\quad \bar{\Phi}=H\Phi^*.
$$
Then
$$
HC=\bar{\Phi}(e^\mu\partial_\mu\Phi+A\Phi I+m\Phi HI),\quad
HC^*=(\partial_\mu\bar{\Phi}e^\mu-I\bar{\Phi}A-mIH\bar{\Phi})\Phi.
$$
\medskip

Using the formula $\tr(UV-VU)=0$, we get
\be
\tr(I\bar{\Phi}A\Phi-\bar{\Phi}A\Phi I)=0,\quad
\tr(HI\bar{\Phi}\Phi-\bar{\Phi}\Phi HI)=0.
\label{lemma}
\ee
Suppose $\Phi\in\Lambda^\even(\E)$ is a solution of the TDE, then, with the
aid of (\ref{lemma}), we obtain
$$
0=\tr(H(C+C^*))=\tr(\bar{\Phi}e^\mu\partial_\mu\Phi+
\partial_\mu\bar{\Phi}e^\mu\Phi)=\tr(\partial_\mu(\bar{\Phi}e^\mu\Phi))=
\partial_\mu j^\mu,
$$
where
$$
j^\mu=\tr(\bar{\Phi}e^\mu\Phi).
$$
Therefore the vector $j^\mu$ is the conservative current. If we take the
1-form
$$
J=g_{\mu\nu}j^\nu e^\mu=j_\mu e^\mu=\Phi\bar{\Phi},
$$
then the divergence $\partial_\mu j^\mu=0$ can be rewritten in the form
$$
\delta J=0.
$$

Finally let us define the Lagrangian from which the TDE can be derived
$$
Lagr_1=\tr(HCI).
$$
Adding the term that describes the free field $A$ to $Lagr_1$, we obtain
\begin{equation}
Lagr=Lagr_1+\tr(F^2),
\label{1040}
\end{equation}
where $F=dA$ is a 2-form, $\tr(F^2)=-\frac{1}{2}f^{\mu\nu}f_{\mu\nu}$,
$f_{\mu\nu}=\partial_\mu a_\nu-\partial_\nu a_\mu$. Hence the Lagrangian
$Lagr$ depends on the following exterior forms: $\Phi\in\Lambda^\even(\E)$,
$\bar{\Phi}\in\Lambda^\odd(\E)$, $A\in\Lambda^1(\E)$.
Using the variational principle \cite{Bogolubov} we suppose that the
exterior forms $\Phi$ and $\bar{\Phi}$ are independent and as the variational
variables we take the 8 functions which are the coefficients of the exterior form
$\bar{\Phi}$ and the 4 functions which are the coefficients of the exterior form $A$.
The Lagrange-Euler equations with respect to these variables give us the
system of equations, which can be written in the form
\begin{eqnarray}
&&(d-\delta)\Phi+A\Phi I+m\Phi HI=0,\nonumber\\
&&dA=F,\label{1050}\\
&&\delta F=J,\nonumber
\end{eqnarray}
where $J=\Phi\bar{\Phi}=\Phi H\Phi^*$.
This system of equation can also be written in the form
\begin{eqnarray}
&&e^\mu(\partial_\mu\Phi+a_\mu\Phi I)+m\Phi HI=0,\nonumber\\
&&\partial_\mu a_\nu-\partial_\nu a_\mu=f_{\mu\nu},\label{1060}\\
&&\partial_\mu f^{\mu\nu}=j^\nu,\nonumber
\end{eqnarray}
where $j^\nu=\tr(\bar{\Phi}e^\nu\Phi)$,
$f^{\mu\nu}=g^{\mu\lambda}g^{\nu\epsilon}f_{\lambda\epsilon}$.

The Lagrangian (\ref{1040}) and the systems of equations
(\ref{1050}),(\ref{1060}) are invariant under the gauge transformation
with the symmetry group ${\rm U}(1)$
\begin{equation}
\Phi\to\Phi\,\exp(\lambda I),\quad A\to A-d\lambda,
\label{1070}
\end{equation}
where $\lambda=\lambda(x)\in\Lambda^0(\E)$ and
$\exp(\lambda I)=e\cos\lambda+I\sin\lambda$.
And they are also invariant under the global (independent of $x$)
transformation (\ref{1020}) with the symmetry group $\Spin(\E)$.
The gauge invariance (\ref{1070}) expresses the interaction of the electron
(fermion) with an electromagnetic field. The global invariance
(\ref{1020}) leads to unusual (compared to tensors) transformation
properties of bispinors under Lorentz changes of coordinates.

The Dirac equation can be generalized to the (pseudo) Riemannian space
$\V$ using a special technique known as the tetrad formalism. We suppose
that the TDE gives another possibility to describe the electron in the
presence of gravity. Here we must take into account the fact that
in Riemannian space there is no invariance of equations under global
transformations of the form (\ref{1020}). Consequently we must use a
gauge (local) transformation instead of the global transformation
(\ref{1020}). This leads to a new gauge field with the $\Spin(\V)$
symmetry group, which we interpret as the gravitational field. For
details of such an approach see  \cite{Marchuk5}.


\subsection{Other tensor equations.}
Consider the Ivanenko-Landau-K\"ahler equation
\cite{Ivanenko},\cite{Kahler}
\be
\Upsilon\rho + iA\rho+im\rho=0,
\label{1080}
\ee
where $\rho=\rho(x)\in\Lambda_\C(\E)$ and $A=A(x)\in\Lambda^1(\E)$.
We arrive at the TDE (\ref{1010}) by multiplying (\ref{1080}) from the right by
the idempotent
$$
t=\frac{1}{4}(e+H)(e-iI)
$$
and using the relations
$$
t=tH,\quad it=tI.
$$
Similarly, multiplying (\ref{1080}) by
$$
t=\frac{1}{2}(e+H)
$$
and using the relation $t=tH$, we arrive at the following equation:
\be
\Upsilon\eta + iA\eta+im\eta H=0,
\label{1090}
\ee
where $\eta\in\Lambda^\even_\C(\E)$.
In the same way, multiplying (\ref{1080}) by
$$
t=\frac{1}{2}(e-ie^5), \quad (e^5=e^{0123}=e^0\wedge e^1\wedge e^2\wedge e^3=
e^0 e^1 e^2 e^3),
$$
and using the relation $it=te^5$ we arrive at the equation
\be
\Upsilon\omega+A\omega e^5+m\omega e^5=0,
\label{1100}
\ee
where $\omega\in\Lambda(\E)$.
Evidently, (\ref{1080}),(\ref{1090}),(\ref{1100}) are tensor equations.


\end{document}